\documentclass[preprint,12pt]{aastex}

\def \eg{{e.g.,}}

\def \etal{{et~al.\null}}
\def \ie{{i.e.,}}

\title{Close Binaries as the Progenitors of the Brightest Planetary 
Nebulae}

\begin{document}

\shorttitle{Close Binaries and Planetary Nebulae}

\author{Robin Ciardullo, Steinn Sigurdsson}
\affil{Department of Astronomy \& Astrophysics, The Pennsylvania State
University \\ 525 Davey Lab, University Park, PA 16802}

\email{rbc@astro.psu.edu, steinn@astro.psu.edu}

\author{John J. Feldmeier\altaffilmark{1}}
\affil{National Optical Astronomy Observatories, P.O. Box 26732, Tucson, AZ 
85726}

\email{johnf@noao.edu}

\and

\author{George H. Jacoby}
\affil{WIYN Observatory, P.O. Box 26732, Tucson, AZ 85726}

\email{jacoby@wiyn.org}

\altaffiltext{1} {NSF Astronomy and Astrophysics Postdoctoral Fellow}

\begin{abstract}
We investigate the possible progenitors of the planetary nebulae (PNs) 
which populate the top 0.5~mag of the [O~III] $\lambda 5007$ planetary nebula 
luminosity function (PNLF{}).  We show that the absolute luminosity of the PNLF 
cutoff demands that the central stars of these most luminous planetaries be 
$\gtrsim 0.6 M_{\odot}$, and that such high-mass PN cores must exist in every
galaxy.  We also use the bolometric-luminosity specific PN number density to 
show that in early-type galaxies, [O~III]-bright planetaries are relatively 
rare, with only $\sim 10\%$ of stars evolving to these bright magnitudes.  
We demonstrate that the combination of these two facts implies that either 
all early-type systems contain a small, smoothly distributed component of young 
($\lesssim 1$~Gyr old) stars, or another mechanism exists for creating 
high-core mass planetaries.  We argue that binary-star evolution is this 
second mechanism, and demonstrate that blue stragglers have the appropriate 
core properties and number density to explain the observations.  We discuss the 
implications of this alternative mode of stellar evolution, and speculate on 
how coalesced binaries might affect the use of PNs for measuring a galaxy's
star-formation history and chemical evolution.

\end{abstract}


\section{Introduction}
The bright-end cutoff of the [O~III] $\lambda 5007$ planetary nebula
luminosity function (PNLF) is one of the most important standard candles on
the extragalactic distance ladder.  The technique is well-tested, extremely
accurate ($\lesssim 10\%$), and applicable throughout the Local Supercluster 
\citep{mudville, chile}.  In fact, since the PNLF technique is equally 
effective for both spiral and elliptical galaxies, its distances provide
a crucial link between the Population~I and Population~II distance scales.

Unfortunately, the PNLF method has one significant failing.  Despite several
attempts at modeling the function, there is still no robust theory for the
physics of the phenomenon.  Indeed, the simple argument of \citet{jacoby96}
and the detailed simulations of \citet{mendez} and \citet{marigo} 
agree that, once the age of a stellar population exceeds $\sim 1$~Gyr, 
the PNLF cutoff should fade dramatically with age.  This prediction is in 
direct conflict with the numerous internal and external tests that have been 
applied to the method \citep{mudville, p12, chile}.

The studies of \citet{mendez} and \citet{marigo}, along with those of others
\citep[\eg][]{p1, han} have focussed exclusively on understanding how the
absolute magnitude of the PNLF cutoff varies with population age and
metallicity.  What these studies have not considered is how the production
rate of [O~III]-bright planetaries evolves with time.  This is an important
omission, since the number of PNs populating the top part of the luminosity 
function provides an important constraint on the progenitors of these
extremely luminous objects.

In this paper, we discuss the possible progenitors of PNs in the top half
magnitude of the [O~III] $\lambda 5007$ planetary nebula luminosity function.
In \S 2, we review the three observational constraints which any successful 
model must satisfy: the invariance of the PNLF cutoff to population age and
metallicity, the absolute brightness of the cutoff, and the fraction of a 
galaxy's stars which evolve through this extremely luminous stage.  In \S 3, 
we use these constraints to identify the likely progenitors of [O~III]-bright 
PNs.  We suggest that, although single stars with initial masses of 
$M \gtrsim 2 M_{\odot}$ can provide many or most of a galaxy's 
brightest planetaries, a non-negligible fraction of the PNs in the top 
$\sim 0.5$~mag of the luminosity function come from coalesced binary 
stars.  In particular, we identify blue stragglers as the likely progenitors 
of [O~III]-bright planetaries, and show that production rate of these objects
is close to that needed to explain the PN counts in elliptical galaxies.  We 
conclude by discussing the implications this hypothesis has for the PNLF of 
spiral galaxies, and how binary star evolution may affect efforts to use 
planetaries to probe galactic evolution.

\section{Constraints on the PNLF}
PNLF measurements have now been made in $\sim 50$~galaxies, ranging from
normal ellipticals and lenticulars, such as NGC~3379, M86, and M87, to
late-type spirals and irregulars, such as M33, M101, and the Large
Magellanic Cloud.  In each system, the [O~III] $\lambda 5007$ PN luminosity
function has a distinctive bright-end cutoff, which can be modeled with 
the truncated exponential
\begin{equation}
N(M) \propto e^{0.307 M} \{ 1 - e^{3 (M^* - M)} \}
\end{equation}
In the above equation, $M$, the absolute magnitude of a planetary, is related 
to monochromatic [O~III] $\lambda 5007$ flux (in ergs~cm$^{-2}$~s$^{-1}$) by
\begin{equation}
M = -2.5 \log F_{5007} - 13.74
\end{equation}
and $M^*$ is the absolute magnitude of the most luminous planetary
\citep{p1,p2}.  Any model which seeks to explain this function must not only
reproduce this shape, but also satisfy three additional criteria.

\subsection{The Invariance of $M^*$}
The fact that the PNLF cutoff does not depend on metallicity is not a
surprise.  From the beginning, it was realized that oxygen acts as a 
thermostat in the nebula, and the (weak) abundance dependence that remains
is almost exactly canceled by opacity effects inside the progenitor star
\citep{p1, lattanzio, djv92}.  Thus, unless the system is extremely metal-poor,
\ie\ [O/H] $\lesssim 8.3$, the metallicity dependence of $M^*$ should be
negligible \citep{p12}.  What was not obvious was why the absolute magnitude of
the cutoff does not depend on age.  The energy which powers an [O~III]-bright
PN comes from its central star, and the luminosity emitted by this central star
is a sensitive function of its mass \citep[$L \propto M^3$;][]{vw94, blocker}.
Since the mass of a post-AGB core depends on the progenitor's
main-sequence mass through the initial mass-final mass relation 
\citep{weidemann, claver}, it is natural to conclude that the absolute 
magnitude of the PNLF cutoff will fade with time.  Yet as numerous 
observations have shown, this dimming does not occur \citep{mudville, p12, 
chile}.

There are a number of processes which can limit the [O~III] flux of 
Population~I planetaries.  The rapid evolutionary times associated with high
mass central stars may not allow the nebulae to reach the low densities
needed for efficient forbidden line cooling \citep{p1, marigo}. 
Alternatively, the short time between the onset of the fast wind 
and ionization may limit the amount of O$^{++}$ available for collisional 
excitation \citep{marigo}.  Finally, as pointed out by \citet{cj99}, the 
[O~III] emission around high mass central stars may be self-extincted by the 
objects' dense, dusty envelopes \citep{cj99}.  Nowhere is this last mechanism 
more evident than in the Large Magellanic Cloud, where, without circumstellar 
extinction, half of the 16 brightest PNs analyzed by \citet{md1, md2} would 
appear brighter than $M^*$.  (In fact, one heavily extincted LMC planetary, 
SMP~89, has an intrinsic [O~III] $\lambda 5007$ flux that exceeds $M^*$ by 
over 0.7~mag!)

The above processes all serve to place an upper limit on the [O~III] flux
emitted by a galaxy's brightest planetary nebulae.  Unfortunately, there is
no equivalent set of conditions which places a lower limit on this quantity.
This leads to a second constraint.

\subsection{The Absolute Value of $M^*$}
Observations of planetary nebulae in galaxies with well-determined Cepheid 
distances yield a value for the PNLF cutoff magnitude of $M^* = -4.47 \pm 0.05$
\citep{p12}.  In other words, planetary nebulae at the bright end of the 
[O~III] luminosity function emit more than $600 \, L_{\odot}$ of power at
5007~\AA.   Note, however, that this is only a small fraction of the object's 
total energy:  both models and observations indicate that no more than
10\% of a central star's total luminosity comes out in this line 
\citep[\eg][]{p1, jc99, marigo}.   Consequently, the post-AGB stars which
power a galaxy's brightest planetaries must, at a minimum, have luminosities
in excess of $\sim 6000 L_{\odot}$.  Indeed, in the bulge of M31, 3 out of the
12 planetaries analyzed by \citet{jc99} have central star luminosities 
brighter than $14,000 L_{\odot}$.

Such high luminosities are a major problem for PNLF models.   In order to 
generate $\sim 6000 L_{\odot}$ of power, a PN central star must have a mass of 
at least $\sim 0.6 M_{\odot}$ \citep{vw94, blocker}.  Moreover, as 
Table~\ref{tab1} suggests, the actual lower mass limit is probably larger.
In the table, we have listed those SMC, LMC, and M31 PNs with absolute [O~III]
$\lambda 5007$ magnitudes within 0.5~mag of $M^*$, and with central stars
whose properties have been derived via a spectrophotometric analysis.
Except for the SMC PNs, whose core luminosities, masses, and $M^*$ are 
suppressed by low metallicity \citep{djv92, p12}, the Local 
Group's [O~III]-bright planetaries all have cores greater than 
$M \sim 0.62 M_{\odot}$.  This mass is significantly larger
than the $0.56 M_{\odot}$ value which is typical for white dwarfs in the
solar neighborhood \citep{madej}.  More importantly, the initial mass-final
mass relation for solar metallicity stars predicts that the progenitors of
$0.62 M_{\odot}$ cores should have main-sequence masses close to 
$\sim 2.2 M_{\odot}$ \citep{weidemann, claver}.  Such high mass objects are 
not normally associated with early-type galaxies, and in M32, which is
considered to be a ``young'' elliptical \citep[\eg][]{trager}, they are 
definitely not present \citep{brown}.  This situation is even worse when
one considers the effect of metallicity:  if the models of \citet{vw93} and
\citet{frantsman} are correct, then the main sequence masses of [O~III]-bright 
PNs in the metal-rich environments of giant ellipticals need to be even 
greater.

\subsection{The Number of [O~III]-Bright Planetary Nebulae}
The third constraint on the PNLF comes from the number of PNs which
populate the bright-end of the luminosity function.  If we let $\alpha$
be the observed number of planetary nebulae per unit (bolometric)
luminosity of the underlying parent stellar population, then
\begin{equation}
\alpha = B \, t \, f
\end{equation}
where $B$ is the population's luminosity specific stellar evolutionary flux, 
$t$, the mean time that PNs spend in the [O~III]-bright stage, and $f$, the 
fraction of stars turning off the main sequence that evolve into [O~III]-bright
planetaries.  For old stellar populations, such as those found in elliptical
and lenticular galaxies, 
$B \sim 2 \times 10^{-11}$~stars~yr$^{-1}~L_{\odot}^{-1}$, 
independent of age, metallicity, or the initial mass function \citep{renzini}.
Since the expected PN lifetime can be estimated from post-AGB stellar 
evolutionary tracks \citep{vw94, blocker} and models of nebular evolution 
\citep{mendez, djv92, marigo}, measurements of $\alpha$ can be used to place 
a strong constraint on the percentage of stars which contribute to the PNLF 
cutoff.

Estimates of $\alpha$ currently exist for over 20 early-type galaxies.  
Unfortunately,  most of these data are in terms of $\alpha_{2.5}$, the 
normalized number of planetaries within 2.5~mag of $M^*$.  Not only does such 
a definition risk including fainter PNs which never reached the bright end 
of the luminosity function, but it is also based on an 
extrapolation.  Because PN surveys outside the Local Group typically extend 
less than $\sim 1$~mag down the luminosity function, $\alpha_{2.5}$ is usually
derived by assuming that equation (1) is valid at all magnitudes.
This may not be the case.  While an exponential extrapolation appears adequate
for some older populations, such as that found in the bulge of M31,
star-forming systems have a PNLF that turns over $\sim 2$~mag below the 
cutoff \citep{jd02, m33}.   To avoid any assumption about the form
of the luminosity function, we therefore restrict our analysis to
those objects at the very top of the luminosity function.

Table~\ref{tab2} contains measurements of $\alpha_{0.5}$, the number of 
planetary nebulae observed within 0.5~mag of $M^*$, normalized to the amount 
of bolometric luminosity surveyed.  These numbers were derived from the studies
listed in the table, using updated values for $M^*$ and Galactic reddening 
\citep{p12, schlegel}, and the techniques described in the original papers. 
Specifically, each value of $\alpha_{0.5}$ was obtained by fitting the 
observed PNLF to the empirical function given in equation (1), and then 
normalizing the predicted number of PNs in the top half magnitude of the 
luminosity function to the total amount of bolometric luminosity sampled 
in the survey region.

The errors quoted in Table~\ref{tab2} represent only the formal uncertainties 
of the fitting procedure, and do not include errors in the luminosity 
normalizations.  This additional term is likely to be small, $\sim 10\%$.  The
greatest uncertainty in the normalizations probably comes from the bolometric
corrections, which were computed from the galaxies' spectral energy 
distributions, as defined by broadband optical and infrared colors.   Since 
these photometric measurements are not always homogeneous or complete, errors
of up to $\sim 0.1$~mag are possible.  Fortunately in all cases, the derived 
bolometric corrections are within 0.1~mag of those predicted for old stellar 
systems by population synthesis models \citep[\ie][]{worthey}.

Figure~\ref{fig1} shows the behavior of $\alpha_{0.5}$ as a function of
several galactic parameters.  For two of the variables, the globular cluster
specific frequency \citep{kisslerpatig, rhode} and the [MgFe] composite
absorption line index \citep{gonzalez, tantalo}, $\alpha_{0.5}$ shows no 
strong trend.  However, the other four plots show correlations which are 
significant at greater than the 98\% confidence level.  Galaxies with
absolute magnitudes fainter than $M_B \sim -19$ and colors bluer than
$(V-I) < 1.1$ have $\sim 1$~[O~III]-bright PN for every 
$4 \times 10^8 L_{\odot}$ of bolometric light.  In larger and redder galaxies, 
the luminosity specific number of PNs declines as the galaxy's UV upturn, as 
measured by the {\sl IUE\/} satellite \citep{burstein}, increases.  This effect 
is dramatic, so that in the most extreme cases, $\alpha_{0.5}$ is almost an 
order of magnitude smaller than in low-luminosity objects.  

The last panel of Figure~\ref{fig1} shows the behavior of $\alpha_{0.5}$ as a
function of the strength of the galaxy's H$\beta$ absorption line.  Although
the equivalent width of this feature depends on a number of factors, such as
the system's metallicity and the $\alpha$-element to iron ratio, the index
is generally considered to be a good indicator of population age 
\citep[\eg][]{gonzalez, worthey, tantalo}.  Thus, the direct correlation
between $\alpha_{0.5}$ and H$\beta$ strength suggests that younger 
populations are more efficient at producing [O~III]-bright planetaries
than older, lower-turnoff mass systems.

To translate $\alpha$ into stellar fractions, we need an estimate of the time
that a typical PN remains in the top 0.5~mag of luminosity function.  This is 
not difficult to derive.  If $\sim 10\%$ of a PN's energy comes out in 
[O~III] $\lambda 5007$, then the exciting star of an object in this stage of
evolution must have $L > 3900 L_{\odot}$ and $T \gtrsim 50,000$~K{}.  Using
these numbers, one can simply use stellar evolutionary tracks \citep[\ie][]
{vw94, blocker} to estimate how long a particular central star satisfies these
conditions.   Alternatively, one can take the approach of \citet{marigo} 
and place evolving post-AGB stars within dynamically evolving nebular models, 
and track the evolution of all the PN's emission lines as a function of time.  
Such an analysis shows that, although the precise timescale for [O~III] 
evolution depends on a number of parameters, such as central star mass, 
post-AGB transition time, and the composition of the energy producing shell, 
a value of $t \sim 500$~yr is appropriate for the brightest PNs in a galaxy.

The right hand axes of Figure~\ref{fig1} translate $\alpha$ to $f$ using
the relationship given in equation (3) and a lifetime for [O~III]-bright PNs
of 500~yr.  The absolute numbers are intriguing: despite the scatter, and the
uncertainties inherent in our estimation of $t$, it is clear that
in early-type systems, only a small fraction of the stars ever reach the
tip of the [O~III] $\lambda 5007$ luminosity function.  In small galaxies,
and galaxies with recent star formation, $\sim 25\%$ of the stars which
turn off the main sequence eventually evolve into planetaries which populate 
the bright end of the PNLF{}.  In redder systems, this fraction drops to only
$\sim 5\%$.  These small values provide an important clue for understanding the 
planetary nebula luminosity function.

\section{The Progenitors of the Brightest PNs}
To summarize the above section:  the progenitors of the PNs which populate 
the top 0.5~mag of the PNLF a) have core masses of $\gtrsim 0.6 M_{\odot}$, 
b) are present in all galaxies, and c) are rarer in redder populations than in 
systems which may have undergone recent star formation.  Obviously, the 
greatest challenge to satisfying these conditions lies in reconciling the need 
for high mass central stars with the observed colors and spectral line 
indices of elliptical and lenticular galaxies.

In principle, there are two ways to generate high-core mass planetaries
in early-type systems.  The first is through the normal
evolution of a trace population of stars with turnoff masses greater than
$\sim 2 M_{\odot}$.  If, as Figure~\ref{fig1} demonstrates, intermediate
mass objects contribute between $\sim 5\%$ and $\sim 25\%$ of an elliptical
galaxy's total luminosity, then all the constraints provided by the PN 
observations are satisfied.  Unfortunately, there is good reason to believe 
that this is not the case.   Turnoff masses of $\sim 2 M_{\odot}$ belong 
to populations with ages of $\sim 1$~Gyr \citep{ibenlaughlin}.  Components
this young should be detectable via their effect on a galaxy's colors and 
spectral line indices, even if they comprise only $\sim 5\%$ of the 
system's total light \citep{worthey, tantalo}.  More importantly, as 
Figure~\ref{fig1} indicates, every early-type galaxy would have to have such 
a component.  While it is possible that some objects (such as NGC~1316 and 
4382) have acquired an intermediate age population via a recent accretion or
merger \citep{goudfrooij, terlevich}, this explanation cannot work for all 
objects.   Galaxies, such as the ``standard'' elliptical NGC~3379 
\citep{devauc} and the bulge of M31 display no evidence of interaction or 
star formation in the recent past \citep{trager}.

A second, more plausible, explanation for the PNLF cutoff invokes an 
alternative form of stellar evolution.  In the solar neighborhood, roughly 
two-thirds of the stars are binaries, and a large fraction of these systems 
interact at some time during their evolution \citep{dq91}.  Those interactions 
which occur on the giant or asymptotic giant branch [the Case~B and C scenarios
of \citet{kippenhahn} and \citet{lauterborn}] usually result in a dynamic 
instability and the ejection of the giant star's envelope 
\citep[\eg][]{ibenlivio}.  However, if the primary is still in its 
core hydrogen burning phase when Roche-lobe overflow occurs 
(\ie\ Case A), then the mass transfer is likely to be conservative, and the 
result can be a star whose mass is greater than that appropriate for its age.
Indeed, there is ample evidence to suggest that this \citet{mccrea} style mass 
transfer is at least partially responsible for the field population of blue 
straggler stars \citep{mateo, preston, carney}.  

Can Case~A mass transfer alone explain the PNLF cutoff?  To answer this
question, we begin by tentatively identifying blue stragglers as the possible 
progenitors of the [O~III]-bright planetaries of elliptical galaxies.   Such an
identification is not unreasonable.   Most blue stragglers have masses that
are less than twice that of their population's turnoff mass \citep{shara}.
For early-type systems with turnoffs of $\sim 1 M_{\odot}$, this places
the stars in the mass range needed to produce the high-mass cores
associated with the PNLF cutoff.   Moreover, in the process of coalescence, 
blue stragglers should acquire a significant amount of angular momentum.
This added rotational support has the potential to prolong the stars'
lives, and allow them to increase their core masses over and above that 
predicted by the single star initial mass-final mass relation.   Whether this
actually happens is still an open question:  although core growth is 
pronounced in stars with initial masses $\gtrsim 5 M_{\odot}$ \citep{doming},
the longer lifetimes of lower mass objects may enable the stars to shed
much of their angular momentum while still close to the main sequence
\citep{sills}.  Yet, even if PN core masses are not enhanced by the 
merger process, binary evolution can still allow populations as old as 
$\sim 10$~Gyr to make [O~III]-bright planetaries.  The scenario therefore
provides a natural explanation for the existence of $M^*$ planetaries in
elliptical galaxies.

If blue stragglers are the antecedents of PNs near the top of the luminosity
function, then the number of planetaries, relative to blue stragglers, should
simply be the ratios of the objects' lifetimes.  As detailed in the preceding
section, the lifetime of PNs in the top 0.5~mag of the luminosity function
is $\sim 500$~yr.  To estimate the lifetime of blue stragglers, we can use the 
results of \citet{lombardi}, who found that the product of an oblique merger 
between two $0.8 M_{\odot}$ stars spends roughly $\sim 750$~Myr on the main 
sequence.  Since the progenitors of [O~III]-bright PNs are likely to be 
slightly more massive than this, a value of $\sim 500$~Myr seems reasonable
for our analysis.  Based on this number, old stellar populations should have 
$\sim 10^6$ more blue stragglers than [O~III]-bright planetary nebulae.

To test this prediction, we first examine the stellar content of Galactic
globular clusters.  This is a difficult place to investigate blue straggler
densities, since collisions offer a separate channel for the objects'
creation and destruction \citep[\eg][]{davies, mapelli}.  Nevertheless, 
the environment can serve as a benchmark, since reasonably complete blue 
straggler surveys now exist for 56 Galactic globulars \citep{piotto04}.  
The analysis of this database demonstrates that, very roughly, globular 
clusters possess $\sim 10^{-3}$~blue stragglers per unit solar $V$-band 
luminosity.  Since the bolometric correction for old, metal poor systems 
is small \citep[$\sim -0.4$;][]{worthey}, this number is directly 
comparable to $\alpha_{0.5}$, the bolometric luminosity specific density
of planetaries.   The ratio of the two specific frequencies, $\sim 10^6$, is
amazingly close to that inferred from the objects' lifetimes.  This 
agreement presents strong circumstantial evidence that the two evolutionary 
states are connected.

As noted above, globular clusters are not the ideal location for studying
binary-evolved blue stragglers.  In order to investigate the systematics of
these objects, one really needs to survey an environment where stellar 
collisions are negligible.  At present, data exist for only two such 
locations: the Ursa Minor dwarf galaxy \citep{carrera} and the Milky Way's own 
halo \citep{preston}.  Although neither dataset is normalized to population 
luminosity, we can use the ratio of blue stragglers to horizontal branch stars  
as a proxy.   In the Ursa Minor dwarf galaxy, this ratio is $\sim 1.8$;
in the Milky Way's halo, the value is close to $\sim 4$.  For comparison,
in the globular clusters studied by \citet{piotto04}, there are $\sim 0.3$
blue stragglers for every horizontal branch star.  Whether the lower number
for globulars is related to the systems' history of blue straggler formation 
\citep[\eg][]{davies} or some other mechanism is not known at this time.  
However, the data do confirm that the production rate of blue stragglers in 
old stellar populations is close to that needed to explain the PNs in the 
top half magnitude of the PNLF.

\section{Modeling the Binary Star Contribution}
If the above scenario is correct, then there are two separate mechanisms for
populating the top 0.5~mag of the PN luminosity function:  the direct channel,
via the evolution of single stars with initial masses of 
$M \gtrsim 2 M_{\odot}$, and the binary evolution route, where Case~A 
accretion and coalescence creates blue stragglers at or near this mass.  
Obviously, in elliptical galaxies, the latter channel will dominate.  The 
production rate of binary-evolved PNs should fall with time, as fewer and fewer 
blue stragglers attain the mass necessary to produce a 
$\sim 0.6 M_{\odot}$~core.  (There is evidence for this in Figure~\ref{fig1}, 
as the large, red galaxies with weak H$\beta$ indices have systematically fewer
PNs than their bluer counterparts.)  However, based on the evidence from 
H$\beta$ and other absorption line spectral indices \citep{worthey, tantalo},
the single-star evolutionary route contributes almost nothing
to the PNLF cutoff in Population~II systems.

The situation is less clear in galaxies with recent star formation.  Although
the single-star channel for PN production is efficient, only those
stars which formed between 0.1 and 1~Gyr ago (\ie\ the evolutionary times
for stars in the $\sim 2$ to $\sim 5 M_{\odot}$ range) can contribute to the
top $\sim 0.5$~mag of the PNLF{}.  Conversely, though only a small fraction 
of binaries will evolve into [O~III]-bright PNs, this channel is open to
objects as old as $\sim 10$~Gyr (or older, if blue straggler cores are 
enhanced by rotational support).  Thus, it is conceivable that the latter
mechanism is important even in late-type systems.

We can explore this possibility by making a simple model of galaxy evolution
that includes both single stars and close binary systems.  We begin with
the premise that the timescale for the creation of a blue straggler is, in 
some way, tied to the evolutionary timescale of the system's primary star.  
This assumption is problematic: the processes which govern a stellar
coalescence may instead be controlled by the rate at which a magnetized stellar
wind carries away angular momentum \citep{stepien}.  Nevertheless, for the 
purpose of this exploratory calculation, the hypothesis should be acceptable. 
Next, we adopt a model of binary formation in which primaries (and single 
stars) obey a \citet{kroupa} initial mass function, secondaries are drawn from 
the distribution of mass ratios given by \citet{hogeveen}, and the initial log 
period distribution of binaries (in days) is flat between 0 and 6 \citep{abt, 
dq91}.  We then divide our binary stars into three classes: short period 
($P < 10$~days) binaries, which undergo conservative mass transfer and 
coalescence, intermediate period ($10 < P ({\rm days}) < 10,000$) systems, 
which interact on the giant branch and do not form high-mass cores, and long 
period ($P > 10,000$~days) wide binaries, which evolve as single stars.  
Finally, we assume that initially, binary systems outnumber single stars by a 
ratio of two to one \citep{dq91} and that the initial mass-final mass relation 
of \citet{weidemann} and \citet{vw94} applies to both single stars and 
coalesced binaries.  Using these assumptions, we can compute the mass spectrum 
of PN central stars as a function of a system's star formation rate history, 
and determine the fraction of [O~III]-bright planetaries that evolve from close 
binaries.

Figure~\ref{fig2} displays the results of our model for the case of a 
galaxy which has been forming stars at a constant rate for 13~Gyr.  The
figure demonstrates that blue-straggler evolution is important:  almost
20\% of PNs with core masses $M_{\rm core} > 0.6 M_{\odot}$ are evolved from
this alternative form of evolution.  Furthermore, by using the same initial
mass-final mass relation for single stars and coalesced binaries, the model
is ignoring the effects of angular momentum.  If blue straggler cores are
enhanced by as little as 10\% by the rotational support acquired in the
merger process, then populations as old as $\sim 13$~Gyr will be able produce 
[O~III]-bright PNs, and almost 40\% of the brightest planetaries in a 
star-forming galaxy will have binary stars as their progenitors.  

Figure~\ref{fig3} expands on this point by showing how the importance of 
binary evolution changes as a function of a galaxy's history of star formation.
In the figure, we have assumed that the galaxy's star-formation rate has
declined exponentially with time; the abscissa of the figure plots this
timescale, $\tau$.  (For reference, the E0, Sb, Sbc, and Sc galaxies of
\citet{bruzual93} have timescales of $\tau = 1, 2, 4$, and 7~Gyr, 
respectively.)  The ordinate of the plot shows the fraction of PNs with core 
masses greater than $0.6 M_{\odot}$ that are evolved from coalesced binaries. 
The figure confirms that binary-evolved objects are important for defining the 
top end of the PNLF, even in galaxies with intermediate age populations
such as the Magellanic Clouds \citep{olszewski}.  In earlier-type systems with 
little recent star-formation, virtually all of the [O~III]-bright PNs come 
from coalesced binaries.

\section{Implications}
If the above scenario is correct, then it presents a major problem for
the study of PNs within our own Galaxy.  According to Figure~\ref{fig3},
the Milky Way contains two types of high-core mass planetaries,
one descended from intermediate-mass single stars of the recent past,
and one evolved from the lower-mass close binaries formed throughout history.
The two objects may not be easily distinguishable.  If the dynamical coupling
between a progenitor's core and envelope is weak \citep{pinsonneault},
then it may be possible to identify blue-straggler evolved objects via the 
rapid rotation of their PNs' central stars.  Unfortunately, though there is 
good reason to believe that stars retain some of their angular momenta 
through the first ascent of the giant branch \citep[\eg][]{peterson, behr}, 
this rotation may not be able to survive the mass loss on the AGB{}.  On the
other hand, if a blue straggler core sheds its angular momentum via an AGB 
superwind, the result could be a bipolar nebulae.  This does not seem to be
the case, since bipolar planetaries in the Milky Way appear to have a smaller
Galactic scale height (and therefore a younger progenitor age) than their 
elliptical and circular counterparts \citep{corradi, phillips_a}.  
Nevertheless, incompletenesses in the PN sample \citep{phillips_b}, as well 
as our general inability to determine the absolute magnitudes of Galactic
PNs \citep[\eg][]{bensby, phillips04} leave some room for doubt.

The blue straggler hypothesis has several implications for PN identifications
in distant galaxies.  The first concerns the ``dip'' in the luminosity
function that occurs in star-forming systems.  \citet{m33} have argued
that a PNLF formed entirely from high mass planetaries must have a 
deficit of objects between $\sim 2$ and $\sim 4$~mag below $M^*$, due to the 
rapid fading of central stars whose nuclear reactions have recently stopped.  
Indeed, the PNLFs of the SMC and M33 show just this effect:  in these 
star-forming galaxies, the number of intermediate brightness planetaries drops 
significantly \citep{jd02, m33}.  No such deficit is present in the old stellar 
population of M31's bulge \citep{m33}.  In this system, it is likely that 
most of the PNs do not participate in the PNLF cutoff, and join the luminosity 
function at fainter magnitudes.  These lower mass objects fill in the 
deficit produced by the high core-mass, binary-evolved PNs, and cause the PNLF
to appear strictly monotonic.

Another implication of our blue straggler theory involves the stellar
populations of elliptical galaxies.  Under our conservative mass transfer
binary coalescence model, the PNLF of early-type systems is defined by 
binaries whose combined mass on the main sequence is greater 
$\sim 2 M_{\odot}$.  As a population ages, fewer and fewer binaries will 
satisfy this condition; consequently the number of such systems, as 
reflected by the $\alpha_{0.5}$, should decrease with time.  The parameter
therefore has the potential to probe a population's turnoff mass, even when 
that turnoff is too faint for direct observation.  Moreover, unlike the results
of integrated light spectroscopy, the information provided by $\alpha_{0.5}$ 
will not be luminosity weighted.  It may therefore be an effective complement
to the traditional approach of studying elliptical galaxy populations via 
absorption-line indices.  Of course, if a population is old enough, no 
binary system will have enough mass to evolve into an $M^*$ planetary, and 
the PNLF will cease to be a reliable distance indicator.  Fortunately, 
judging from the PNLF measurements to date \citep[\ie][]{mudville, chile},
and the presence of a high core-mass planetary in the Galactic globular 
cluster M15 \citep{alves}, this time has not yet occurred.

Finally, the contribution of binary-evolved PNs to the PNLF has important
implications for the use of planetary nebulae as tracers of a system's
star-formation history and chemical evolution.  For example, by analyzing
the line ratios and line fluxes of a set of PNs at a known distance, 
\citet{dopita} has shown that it is not only possible to measure the objects'
chemistry, but also the masses of the PNs' central stars.  Consequently, with
the aid of an initial-mass final-mass relation \citep[\eg][]{weidemann}, it is 
theoretically possible to use planetaries to trace a galaxy's star formation
and chemical enrichment history back $\sim 10^{10}$~yr.  Unfortunately, the 
method only works if there is a unique relation between the mass of a 
PN's core and the main-sequence mass of its progenitor star.  Our observations
suggest that this is not the case, at least for the brightest PNs in a galaxy.
Because a substantial fraction of these objects evolve from binary stars,
the use of bright planetaries for chemical evolution studies is problematic.

The situation should improve as one observes further down the luminosity
function.  By their very nature, blue stragglers tend to produce PNs
with high-mass central stars.   Their existence will therefore 
complicate PN-based efforts to probe a galaxy's recent ($\sim 2$~Gyr)
star formation history \citep[\eg][]{villaver04}.  However, the inverse
relationship between main-sequence mass and stellar lifetime, coupled 
with the initial mass-final mass relation, guarantees that single stars will 
be the dominant source of lower mass cores.   Spectroscopy of objects
$\gtrsim 2$~mag down the [O~III] $\lambda 5007$ PNLF is therefore still a
viable method of probing the distant past of a stellar population.  
Unfortunately, reaching those depths in the distant galaxies of Leo~I and 
Virgo will be extremely difficult.

This research made use of the NASA Extragalactic Database and was supported
in part by NSF grants AST 00-71238 and AST 03-02030.

\clearpage
\begin{deluxetable}{lccccl}
\tablewidth{0pt}
\rotate
\tabletypesize\small
\tablecaption{Local Group [O III] Bright Planetaries}
\tablehead{
&&\multicolumn{2}{c}{Central Star} & \\
&&\colhead{Luminosity} &\colhead{Mass} 
&\colhead{Percent Energy} \\
\colhead{Name}
&\colhead{($M-M^*)$\tablenotemark{a}}
&\colhead{($L/L_{\odot}$)}
&\colhead{($M/M_{\odot}$)}
&\colhead{Emerging at 5007~\AA}
&\colhead{Source} }
\startdata
M31 PN 1   &0.49 &14,450 &0.80 &2.7\%  &\citet{jc99} \\
LMC SMP 62 &0.03 & 4,710 &0.62 &12.6\% &\citet{dm91b}\tablenotemark{b} \\
LMC SMP 73 &0.28 & 7,120 &0.66 &6.6\%  &\citet{dm91b}\tablenotemark{b} \\
LMC SMP 99 &0.32 & 5,700 &0.63 &8.0\%  &\citet{dm91b}\tablenotemark{b} \\
LMC SMP 78 &0.38 & 6,040 &0.63 &7.1\%  &\citet{dm91b}\tablenotemark{b} \\
LMC SMP 92 &0.42 & 9,130 &0.70 &4.5\%  &\citet{dm91b}\tablenotemark{b} \\
LMC SMP 52 &0.43 &10,000 &0.70 &4.1\%  &\citet{villaver03} \\
LMC SMP 25 &0.46 & 8,900 &$>0.66$\phantom{$>$} &4.5\%  
&\citet{villaver03}\tablenotemark{b} \\
SMC SMP 17 &0.25 & 4,500 &0.59 &9.5\%  &\citet{villaver04} \\ 
SMC SMP 13 &0.41 & 5,240 &0.61 &7.0\%  &\citet{dm91a}\tablenotemark{b} \\

\enddata
\label{tab1}
\tablenotetext{a}{Magnitudes computed from the [O~III] $\lambda 5007$
photometry of \citet{p6} and \citet{p2}, assuming $(m-M)_V = 18.80$ for the
LMC, $(m-M)_V = 19.09$ for the SMC, and $(m-M)_V = 24.67$ for M31.  The
value of $M^*$ in the SMC is assumed to be reduced by 0.25~mag relative to
that of the LMC and M31 by the effects of metallicity \citep[see][]{p12}.}
\tablenotetext{b}{Central star masses estimated by adopting the central
star properties derived by the authors and interpolating in the 
hydrogen-burning post-AGB evolutionary tracks of \citet{vw94}.}

\end{deluxetable}

\begin{deluxetable}{lccl}
\tablewidth{0pt}
\tablecaption{Luminosity Specific PN Densities}
\tablehead{
\colhead{Galaxy} 
&\colhead{$\alpha_{0.5}$}
&\colhead{Uncertainty} 
&\colhead{Source} }
\startdata
NGC 205           &2.27 &$-0.39$/$+0.63$  &\citet{corradi205} \\
NGC 221           &2.43 &$-0.63$/$+1.57$  &\citet{p2} \\
NGC 224 (bulge)   &1.50 &$-0.14$/$+0.15$  &\citet{p2} \\
NGC 1023          &1.65 &$-0.25$/$+0.29$  &\citet{p7} \\
NGC 1316          &0.87 &$-0.12$/$+0.15$  &\citet{p9} \\
NGC 1399          &1.04 &$-0.18$/$+0.25$  &\citet{p9} \\
NGC 1404          &1.14 &$-0.26$/$+0.40$  &\citet{p9} \\
NGC 3031 (bulge)  &1.52 &$-0.15$/$+0.16$  &\citet{p3} \\
NGC 3115          &2.08 &$-0.47$/$+2.08$  &\citet{p12} \\
NGC 3377          &3.02 &$-0.59$/$+0.83$  &\citet{p4} \\
NGC 3379          &1.65 &$-0.23$/$+0.27$  &\citet{p4} \\
NGC 3384          &3.03 &$-0.43$/$+0.54$  &\citet{p4} \\
NGC 4278          &0.80 &$-0.14$/$+0.18$  &\citet{p10} \\
NGC 4374          &1.39 &$-0.22$/$+0.28$  &\citet{p5} \\
NGC 4382          &1.61 &$-0.20$/$+0.23$  &\citet{p5} \\
NGC 4406          &1.08 &$-0.14$/$+0.18$  &\citet{p5} \\
NGC 4472          &0.50 &$-0.09$/$+0.12$  &\citet{p5} \\
NGC 4486          &1.35 &$-0.14$/$+0.14$  &\citet{m87} \\
NGC 4494          &1.17 &$-0.12$/$+0.12$  &\citet{p10} \\
NGC 4594          &1.48 &$-0.12$/$+0.12$  &\citet{n4594} \\
NGC 4649          &0.49 &$-0.12$/$+0.19$  &\citet{p5} \\
NGC 5102          &3.56 &$-0.70$/$+0.92$  &\citet{n5102} \\
NGC 5128          &4.23 &$-0.20$/$+0.20$  &\citet{n5128} \\
\enddata
\label{tab2}
\end{deluxetable}
\clearpage

\figcaption[fig1.eps]{The bolometric-luminosity specific number of PNs
in the top 0.5~mag of the [O~III] planetary nebula luminosity function,
plotted against the parent galaxy's absolute $B$ magnitude, globular
cluster specific frequency ($S$), optical and UV color, and the absorption-line 
indices [MgFe] and H$\beta$.  Circles indicate elliptical galaxies, squares 
are lenticulars, and triangles are spiral bulges; the crosses represent 
systems which contain obvious evidence of recent star formation.  The error
bars represent formal $1 \, \sigma$ uncertainties and are lower limits to the 
true errors.  The right hand axes translate the PN densities to the
fraction of stars turning off the main sequence which evolve into 
[O~III]-bright planetaries.
\label{fig1}
}

\figcaption[fig2.eps]{The expected mass distribution of PN central stars
for a galaxy with a constant star-formation rate.  The dashed line shows
the contribution of single star evolution, while the solid line illustrates
planetaries from both single stars and blue stragglers.  Binary systems which
interact on the red giant or asymptotic giant branch have been ignored.
The mass distribution given by this simple model is consistent
with that observed for Magellanic Cloud planetaries \citep{villaver04} and
suggests that at least $\sim 20\%$ of the high core-mass planetaries of the 
Milky Way are descended from close binary stars.
\label{fig2}
}

\figcaption[fig3.eps]{The fraction of high-core mass planetaries that are 
evolved from close binary stars, as a function of a galaxy's star formation 
rate history.  The abscissa gives the e-folding time of the system's star 
formation; the ordinate shows the fraction of PN cores with $M > 
0.6 M_{\odot}$ that are evolved from binaries.  The solid line assumes that 
the solar metallicity \citet{vw94} initial mass-final mass relation applies
to all objects; the dashed curve shows a model where the core masses of
coalesced binaries have been enhanced by 10\%.  The curves show that
binary-evolved objects contribute to the bright-end of the PNLF in all 
stellar populations.
\label{fig3}
}

\clearpage

\begin{figure}
\figurenum{1}
\plotone{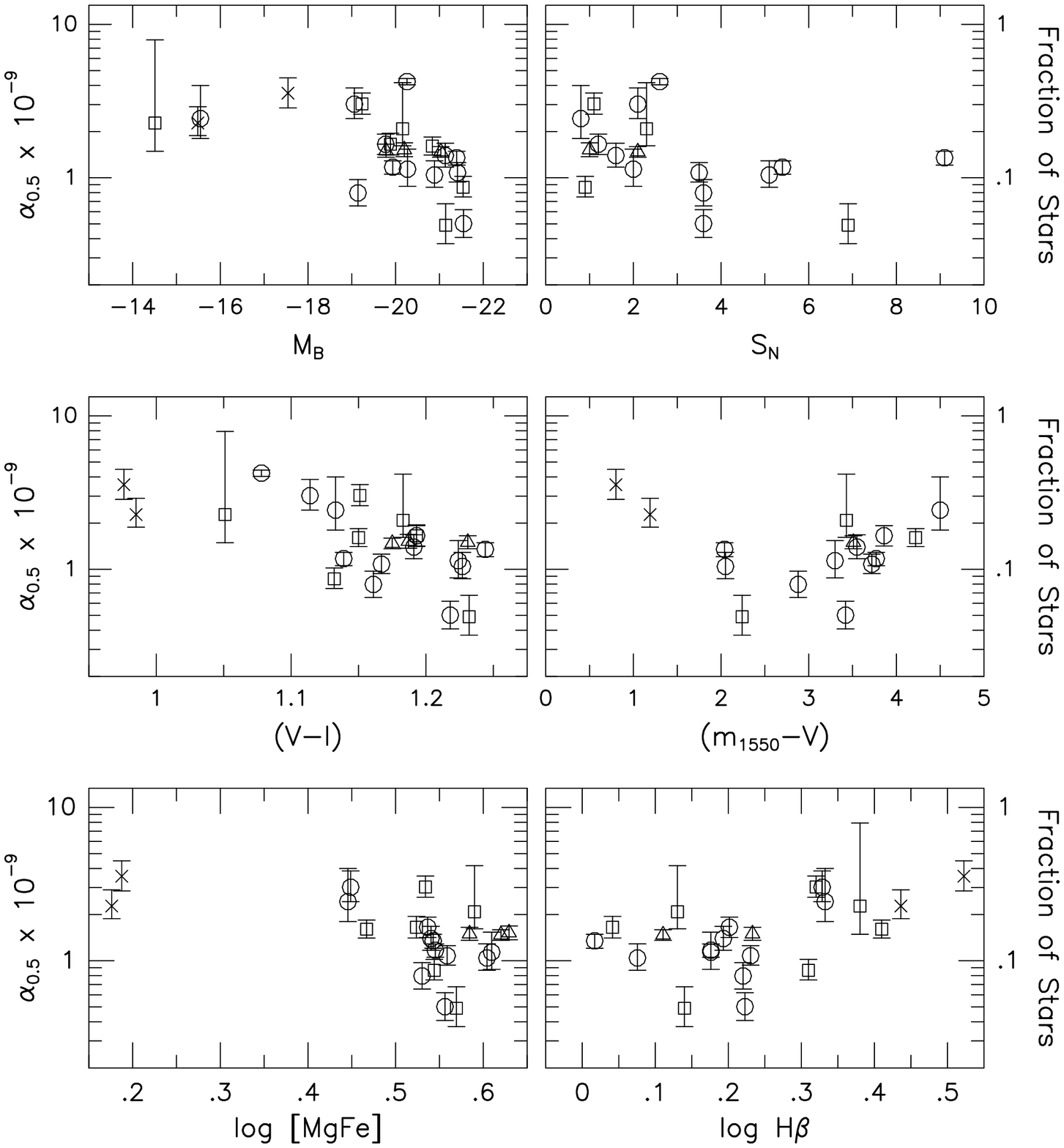}
\end{figure}
\clearpage

\begin{figure}
\figurenum{2}
\plotone{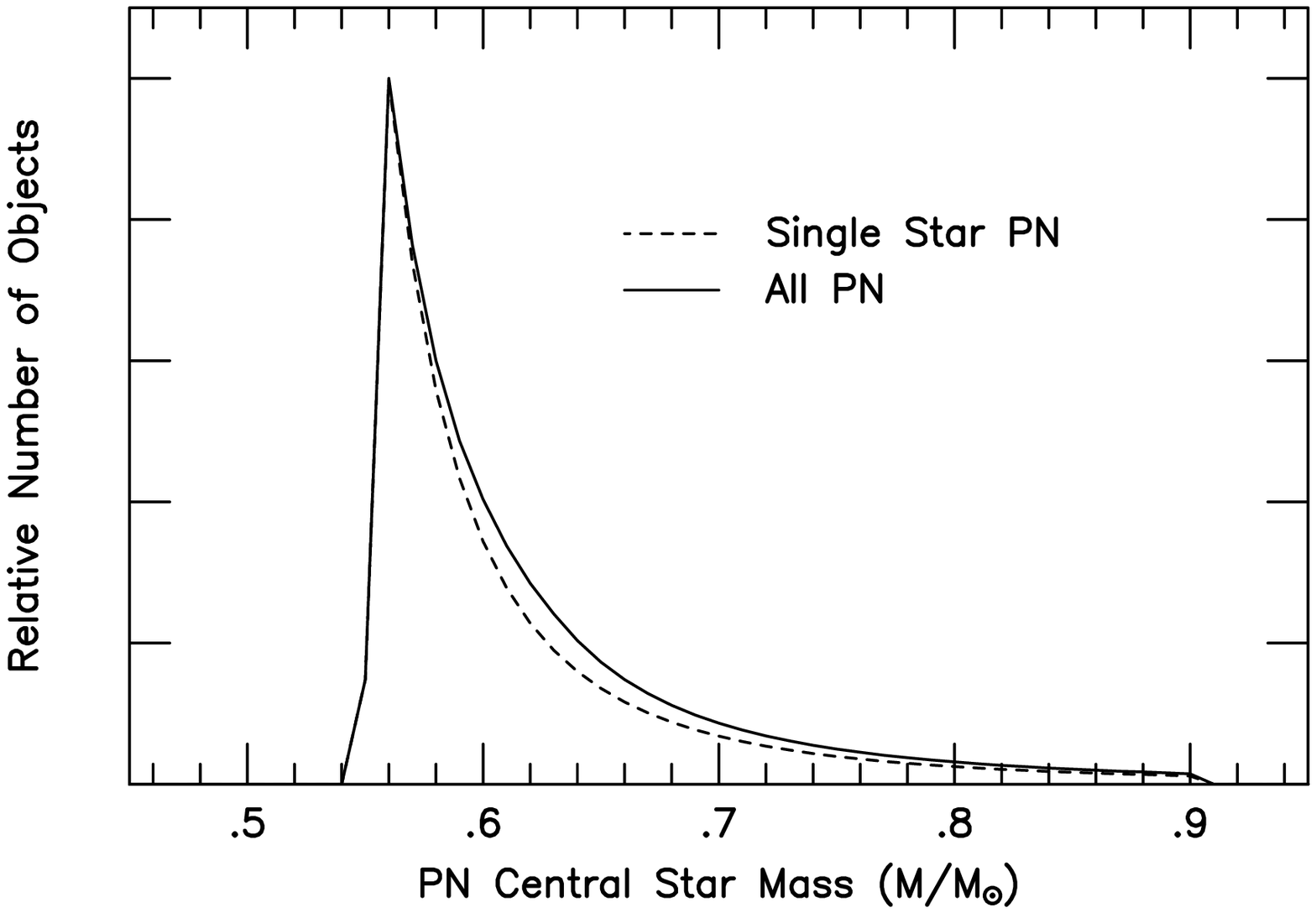}
\end{figure}
\clearpage

\begin{figure}
\figurenum{3}
\plotone{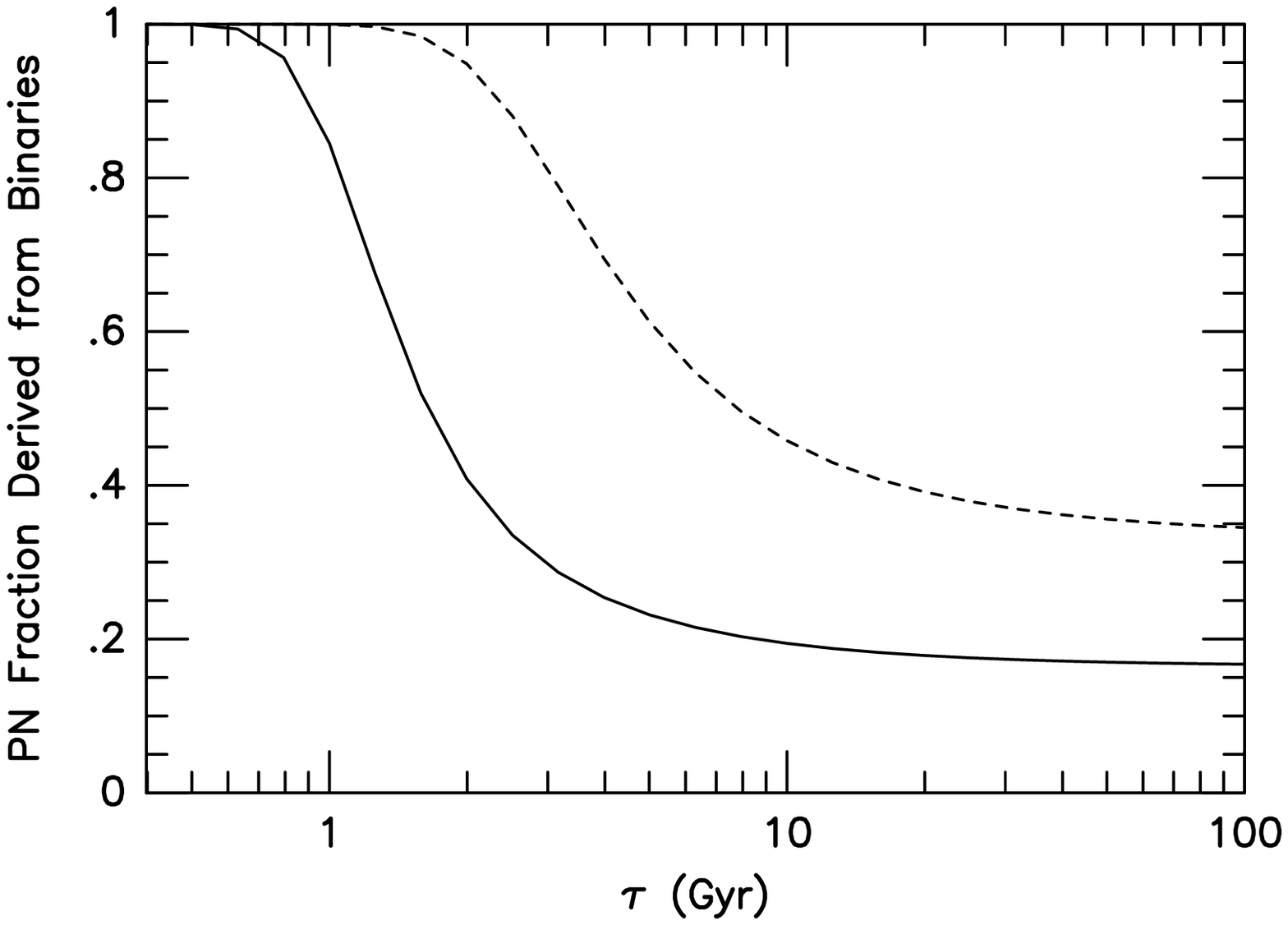}
\end{figure}
\clearpage

\end{document}